\documentclass[aps,prl,twocolumn,showpacs,floatfix]{revtex4}
\usepackage{graphics}
\usepackage{epsfig}
\usepackage{subfigure}
\usepackage{amsmath,amsfonts,amssymb,graphicx}

\begin{document}
\title{Synchronization reveals correlation between oscillators on networks}

\author{Jie Ren$^{1}$}
\author{Huijie Yang$^{2}$}
\affiliation{$^{1}$Department of Physics, University of Fribourg,
CH-1700, Fribourg, Switzerland\\
$^{2}$ Department of Physics and Center for Computational Science
and Engineering, National University of Singapore, Singapore 117542}

\date{today}

\begin{abstract}
The understanding of synchronization ranging from natural to social
systems has driven the interests of scientists from different
disciplines. Here, we have investigated the synchronization dynamics
of the Kuramoto dynamics departing from the fully synchronized
regime. We have got the analytic expression of the dynamical
correlation between pairs of oscillators that reveals the relation
between the network dynamics and the underlying topology. Moreover,
it also reveals the internal structure of networks that can be used
as a new algorithm to detect community structures. Further, we have
proposed a new measure about the synchronization in complex networks
and scrutinize it in small-world and scale-free networks. Our
results indicate that the more heterogeneous and ``smaller" the
network is, the more closely it would be synchronized by the
collective dynamics.
\end{abstract}


\maketitle

The last decade has witnessed the rapid explosion of interest in
complex networks, which are found in many fields as diverse as
biology, technology and social organizations
\cite{review1,review2,review3,review4}. The relation between
structure and function becomes a key area in the study of complex
networks. In the study of networked-dynamics, the emergent
synchronization of interacting oscillators generally has occupied a
privileged position because of its rich applications in variety of
areas ranging from Neuroscience to Sociology
\cite{sync1,sync2,sync3,sync4,sync5,sync6,sync7,sync8,sync9,sync10,sync11,sync12,sync13}.
In neural systems, the brain dynamics is characterized by
synchronization phenomena for a given topology of synapses. And in
metabolic networks, hundreds of interconnected biochemical reactions
are responsible for the biomass and energy fluxes which are adjusted
to optimize the robustness of synchronized behavior. Other example
is the synchronized or coordinated behavior of communication
patterns in social organizations. The study of synchronization
provides us with insights into the key issue: how the collective
dynamics couple the relationships between the systematic function
and the underlying topology.

One of the most successful attempts to understand synchronization
phenomena is due to the Kuramoto model (KM) \cite{KM,KM1}, which is
rich enough for many different contexts, including superconducting
currents in Josephson junction arrays, emerging coherence in
populations of chemical oscillators, and the accuracy of central
circadian pacemakers in insects and vertebrates, \emph{etc.}. This
model describes a population of $N$ coupled phase oscillators which
evolves in time according to the following dynamics
\begin{equation}
\frac{d\theta_i}{dt} = \omega_i + \sigma
\sum_{j}A_{ij}\sin(\theta_{j}-\theta_{i}) + \xi_{i}(t).\label{eq:KM}
\end{equation}
Here, $\theta_i$ represents the phase of the $i$th oscillator,
$\omega_i$ the intrinsic frequency, $\sigma$ the coupling constant,
$A_{ij}$ the effective coupling between the oscillators, and
$\xi_{i}(t)$ is the white noise due to the complicated environment,
with expectation and variance
\begin{eqnarray}
\langle \xi_i(t)\rangle &=& 0, \nonumber \\
\langle \xi_i(t) \xi_j(t') \rangle &=& 2 \delta_{ij} \delta(t-t').
\nonumber
\end{eqnarray}
There are many attractors that each oscillator rotates at its
natural frequency when incoherent. As the coupling strength
increases, some units become resonant. At sufficiently strong
coupling, when coherent, there is only one attractor of the
dynamics, and all oscillators rotate at the same frequency,
$\overline\omega_i$, the average of $\omega_i$. We call it fully
synchronized.

The synchronization problem is solved mainly in the mean-field
approach which, unfortunately, is not usually fulfilled in real
systems. The nontrivial pattern of connectivity in complex networks
brings the non mean-field properties and incorporates many new
questions to the research of synchronization. Because of the elegant
work of Pecora and Carroll \cite{MSF}, the Master Stability Function
(MSF) formalism allows us to analysis the dynamical
synchronizability in terms of purely topological porperties,
independent of details of unit dynamics. However, the MSF requires
several constraints and has limited applications that it only deals
with the stability of the exact synchronized states, \emph{i.e.}, it
is used to study infinitesimal deviations from the synchronization
manifold of chaotic oscillators. However, many realistic dynamics
are only close to or even far from the full synchronization.

In this work we study the KM dynamics departing from the fully
synchronized state. We get the analytic expression of the dynamical
correlation between any two oscillators which reveals the relation
between the network dynamics and the underlying topology. It could
be used as a new algorithm to detect community structures. Moreover,
we propose a new measure about the synchronization in complex
networks and scrutinize it in Small-World and Scale-Free networks.

\begin{figure}
\scalebox{0.75}[0.9]{\includegraphics{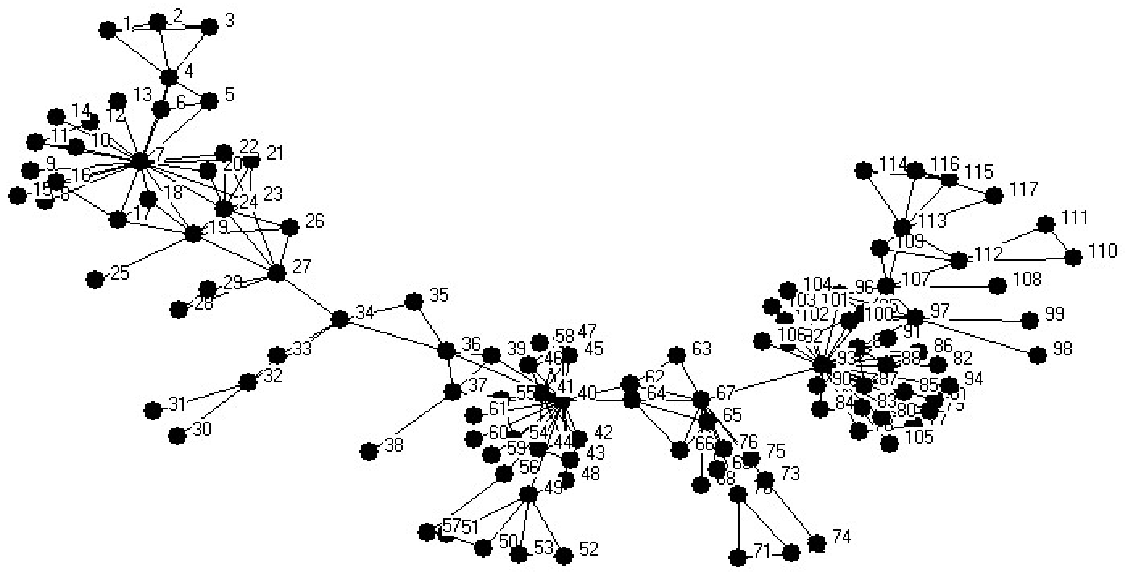}}
\scalebox{0.85}[0.9]{\includegraphics{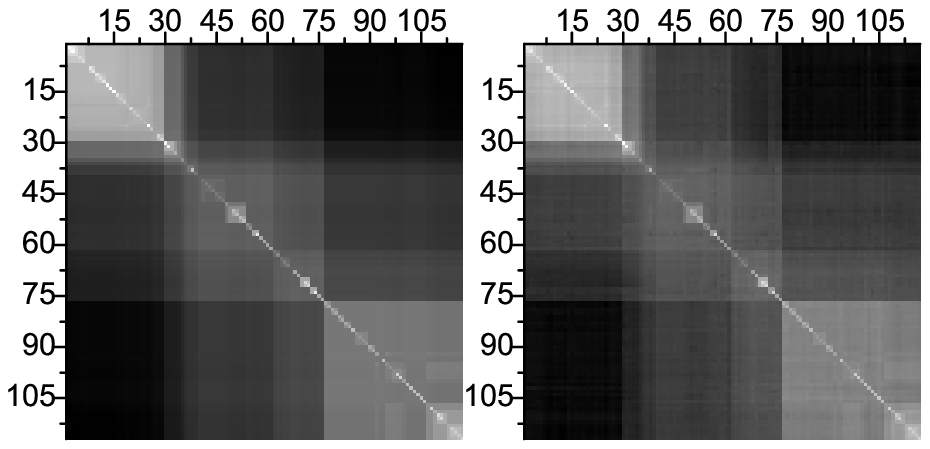}}
\caption{\label{fig:SFI} Top row: Illustration of the Santa Fe
Institute collaboration network. Bottom left: The analytic
correlation matrix resulting from Eq. \ref{eq:cor}. Bottom right:
The simulated results of the correlation between pairs of
oscillators from the Kuramoto model. The colors are a gradation
between white (strong correlation) and black (weak correlation).}
\end{figure}

When the KM dynamics is close to the attractor of synchronization,
the phase differences are small and then the sine coupling function
can be approximated linearly, so that we can analyze the linearized
dynamics of the Kuramoto model in terms of the Laplacian matrix,
\begin{eqnarray}
\frac{d\theta_i}{dt} &=& - \sigma \sum_{j}a_{ij}(\theta_{i}-\theta_{j}) + \xi_{i}(t) \nonumber \\
&=& - \sigma
\sum_{j}L_{ij}\theta_{j} + \xi_{i}(t), \label{eq:Laplace}
\end{eqnarray}
where
$L_{ij}=\delta_{ij}\sum_{m}a_{im}-a_{ij}=\delta_{ij}k_i-a_{ij}$.
$a_{ij}$ stands for the adjacency matrix ($1$ if nodes $i$ and $j$
are connected and $0$ otherwise), $\delta_{ij}$ is the Kronecker
delta and the degree $k_i$ is defined as the number of nodes to
which the node $i$ is connected. Thus, we can rewrite the equation
in terms of the normal modes,
\begin{equation}
\frac{d\vartheta_{\alpha}}{dt} = - \sigma
\lambda_{\alpha}\vartheta_{\alpha} + \zeta_{\alpha}(t),
\end{equation}
where $\vartheta_{\alpha} = \sum_{j}\psi_{\alpha j}\theta_{j}$,
$\zeta_{\alpha} = \sum_{j}\psi_{\alpha j}\xi_{j}$, and $\psi_{\alpha
j}$ denotes the $\alpha$th normalized eigenvector of the Laplacian,
$\lambda_{\alpha}$ is the corresponding eigenvalue for
$\alpha=0,...,N-1$. Considering $\xi_{i}(t)$ is delta correlated, we
can find easily that $\zeta_{\alpha}(t)$ is also delta correlated:
$\langle\zeta_{\alpha}(t)\zeta_{\beta}(t')\rangle=2\delta_{\alpha\beta}\delta(t-t')$.

Without loss of generalization, we set $\sigma=1$,
$\zeta_{\alpha}(0)=0$. Using stochastic calculus methods
\cite{method}, we get
\begin{equation}
\vartheta_{\alpha}(t) = \int_{0}^t e^{- \lambda_{\alpha} (t-t')}
\zeta_{\alpha}(t') dt',
\end{equation}
so that
\begin{eqnarray}
\langle \vartheta_{\alpha}(t)^2 \rangle &=& \int_{0}^t \int_{0}^t
e^{- \lambda_{\alpha} (t-t')} e^{- \lambda_{\alpha} (t-t'')}\langle \zeta_{\alpha}(t') \zeta_{\alpha}(t'') \rangle dt' dt'' \nonumber\\
&=& \int_{0}^t 2 e^{-2 \lambda_{\alpha} (t-t')} dt' \nonumber\\
&=& \frac{1}{ \lambda_{\alpha}}(1-e^{-2 \lambda_{\alpha}t})
\end{eqnarray}
Thus, the time scales of the dynamical relaxation of the eigenmodes
are inversely proportional to the corresponding eigenvalues. And for
large $t\gg 1/2\lambda_{\alpha}$, we have
\begin{equation}
\langle \vartheta_{\alpha}(t)^2 \rangle=1/\lambda_{\alpha}.
\end{equation}
Transforming it back to the basis of $\theta_{i}$ by substituting
$\theta_{i}=\sum_{\alpha}\psi_{\alpha i}\vartheta_{\alpha}$, we find
the steady-state correlation function:
\begin{eqnarray}
\langle (\theta_i-\overline \theta) (\theta_j-\overline
\theta)\rangle &=& \sum_{\alpha=1}^{N-1} \psi_{\alpha i}
\psi_{\alpha j}\langle \vartheta_{\alpha}^2 \rangle \nonumber\\
&=& \sum_{\alpha=1}^{N-1} \frac{1}{\lambda_{\alpha}} \psi_{\alpha i}
\psi_{\alpha j}, \label{eq:cor}
\end{eqnarray}
where $\overline \theta = \frac{1}{N}\sum_i \theta_i$ and
$\langle...\rangle$ stands for the average over initial random
phases. This analytic expression gives the dynamical correlation
between any pair of oscillators just in terms of the properties of
the underlying topology. Actually, the right term of Eq.
\ref{eq:cor} gives the pseudo-inverse of the Laplacian matrix, which
acts like a Green Function. It could be used to detect the community
structure in various complex networks.

One real world example with known community structure is the
collaboration network of the Santa Fe Institute from ref.
\cite{Girvan}. The top row of Fig. 1 illustrates the largest
component of the collaboration graph, which consists of $118$
scientists denoted by nodes. The edge is drawn between a pair of
scientists if they coauthored one or more papers. Obviously, the
scientists of different interests group in different communities.

In the bottom row of Fig. 1, we illustrate the dynamical correlation
between any pairs of nodes on which the KM oscillators is placed.
The left one is the analytic results from the relation Eq.
\ref{eq:cor} and the right is the results of numerical simulations
from Eq. \ref{eq:KM}. They accord with each other very well. In the
two figures we can identify the distinct dynamics-based communities
wherein nodes become strong correlated in groups, coherently with
their topological structure. Moreover, smaller subcommunities can be
found clearly in these dynamics-based communities which corresponds
with the sketch map of the collaboration network. Although some
nodes in the same dynamics-based community are not directly
connected, they are topologically equivalent and belong to the same
structural level. Therefore, they have strong correlation on
dynamics. The dynamics-based groups indicates the different
functional modules. When applied to the specific internal structure,
which might be the fingerprint of different functional groups for
social or biological networks, the dynamical correlation may help us
understand better the interplay between systemic structure and
function.

\begin{figure}
\scalebox{0.8}[0.80]{\includegraphics{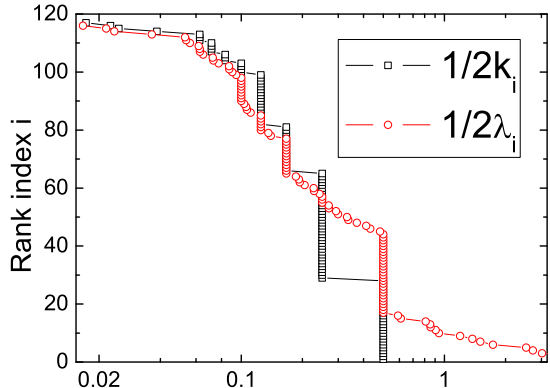}}
\caption{\label{fig:epsart} (color online). Illustration of the same
hierarchical structure in the eigenvalues spectra and degree series.
It implies a progressive cascade of the synchronization pattern from
the central core to the nodes with smaller degrees. The network data
is used as the same as in Fig. \ref{fig:SFI}.}
\end{figure}


In another way, we can use the mean-field approximation to rewrite
the Eq. \ref{eq:Laplace} as:
\begin{eqnarray}
\frac{d\theta_i}{dt} &=& - \sum_{j}A_{ij}(\theta_{i} - \theta_{j}) +
\xi_{i}(t)  \nonumber\\
&=& - k_{i}(\theta_{i} - \overline \theta) + \xi_{i}(t).
\end{eqnarray}
Using the same stochastic calculus method, we get
\begin{eqnarray}
\langle (\theta_i-\overline \theta)^2\rangle = \frac{1}{k_i}(1-e^{-2
k_i t}),
\end{eqnarray}
which indicates that the time scales of the dynamical relaxation of
the observed phases are inversely proportional to the degrees of the
corresponding oscillators. The different connectivities in the
topology give rise to the corresponding order of the eigenvalues,
which are illustrated in Fig. 2. The hubs with the large eigenvalues
induce the fast relaxations, \emph{i.e.}, the evolutional pattern of
synchronization starts first at the hubs and then pervades almost
the whole network in a hierarchical structure across the nodes with
smaller degrees. And for large $t\gg1/2k_i$, we have got $\langle
(\theta_i-\overline \theta_i)^2\rangle = 1/k_i$, which indicates
that hubs are synchronized more closely by the collective dynamics.

For a given realization of networks, we use $R$ to denote the
average variance of phase variables which characterizes the
collective synchronized ability of the underlying network:
\begin{eqnarray}
R = \frac{1}{N} \sum^{N}_{i}\langle
(\theta_i-\overline\theta)^2\rangle ,
\end{eqnarray}
The smaller the value of $R$, the smaller the fluctuation of network
is, \emph{i.e.}, the more synchronized . Substituting the Eq.
\ref{eq:cor} and considering the orthonormality of $\psi_{\alpha
i}$, we have
\begin{eqnarray}
R = \frac{1}{N} \sum_{\alpha=1}^{N-1} \frac{1}{\lambda_{\alpha}}.
\label{eq:sync}
\end{eqnarray}
Therefore, independent of details of unit dynamics, we can
characterize the dynamical synchronizability of networked systems
just in terms of the spectra of the Laplacian matrix, \emph{i.e.},
the purely topological porperties. It is the same as the spirit of
the MSF. However, in contrast to the MSF, the new measure Eq.
\ref{eq:sync} considers the spectrum in its entirety, not only the
extremal eigenvalues. What this new measure emphasizes is on the
fluctuation of the process of synchronization, rather than on
rigorous bounds for the threshold of desynchronization. In the
following, we use this new quantity about synchronization, $R$ to
measure the synchronizable ability in small-world (SW) and
scale-free (SF) networks, respectively.

\begin{figure}
\scalebox{0.8}[0.85]{\includegraphics{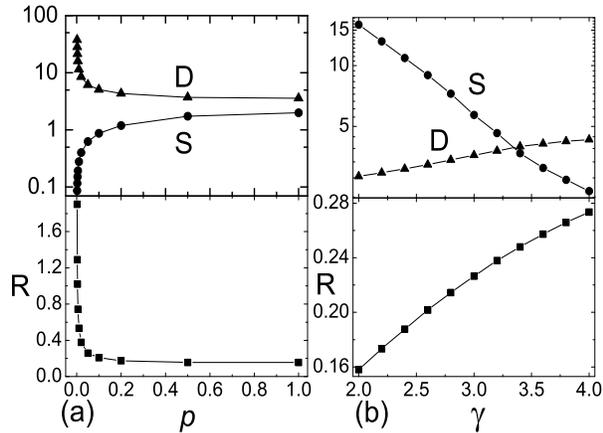}}
\caption{\label{fig:swsf} (a) The synchronizability of SW networks
of size $N=1000$, $\langle k \rangle=8$. The average network
distance $D$ and the new measure $R$ decreases with the rewiring
probability $p$ in the same trend, while the standard deviation $S$
of the degree distribution increases. (b) The synchronizability of
SF networks of size $N=1000$, $k_0=4$. The values of $D$ and $R$
increase with the scaling exponent $\gamma$, while the values of $S$
are decreased. Since less $R$ means more synchronizable, all the
results in SW and SF networks indicate that the network would be
more synchronizable as it becomes ``smaller" and more heterogeneous.
All plots are averaged over 100 realizations.}
\end{figure}

We first consider the Watts-Strogatz model of the SW networks
\cite{WS}, which is constructed by a rewiring process on a regular
ring graph. The probability of rewiring connection is controlled by
a parameter $p$, by tuning which the obtained network possesses both
short average distance and high heterogeneity that nodes no longer
have the same degrees, instead they follow a Poisson distribution.
We define $S$ as the standard deviation of the degree distribution
and $D$ as the average network distance, averaged over all pairs of
nodes. The upper layer of Fig. \ref{fig:swsf}(a) shows the
dependence of $S$ and $D$ on the rewiring probability $p$ in the SW
network. As $p$ increases, the variance $S$ increases, which implies
that the degree distribution becomes more broad and heterogeneous.
And as expect, the network distance $D$ decreases as the rewiring
probability $p$ (or the heterogeneity of the degree distribution)
increases. In the lower layer of Fig. \ref{fig:swsf}(a), $R$ is
observed to decrease, which implies enhancement of the
synchronizable ability.

Secondly, we implement the same measures on the SF network. The
model \cite{SF} is generated by randomly connecting nodes, forcing
each nodes $i$ to have connectivity $k_i\geq k_0$ which follows the
probability distribution $P(k)\sim k^{-\gamma}$. Note that
decreasing $\gamma$ increases $S$ by inducing the longer tail in the
connectivity distribution and smaller $\gamma$ gives rise to shorter
$D$, as exhibited in the upper layer of Fig. \ref{fig:swsf}(b). And
in its lower layer, $R$ is observed to increase as $\gamma$ is
increased, which indicates less synchronized when less
heterogeneous.

The results shown in Fig. \ref{fig:swsf} imply that the
synchronizable ability on the SW and SF networks is improved as the
heterogeneity of the degree distribution is increased or as the
average network distance is decreased. In other words, as the
network becomes more heterogeneous and ``smaller", with $S$ gets
larger and $D$ gets shorter, it would becomes synchronized more
closely by the collective dynamics.

In summary, we have investigated the synchronization dynamics of the
KM departing from the fully synchronized state. We have got the
analytic expression of the dynamical correlation between pairs of
oscillators that reveals the relation between the network dynamics
and the underlying topology. Moreover, it also reveal the internal
structure of networks that can be used as a new algorithm to detect
community structures. Further, we have proposed a new measure about
the synchronization in complex networks and scrutinize it in SW and
SF networks. Our results indicate that the more heterogeneous and
``smaller" the network is, the more closely it would be synchronized
by the collective dynamics. We hope our study can provide new
insights into the understanding of the role of synchronization
between the network structure and function. We also expect it can
provide new tools to detect community structure and to analyze the
ubiquitous synchronization phenomenon.

\end{document}